# Multi-party quantum private comparison of size relation with $d$-level single-particle states

Chong-Qiang Ye, Tian-Yu Ye *

College of Information & Electronic Engineering, Zhejiang Gongshang University, Hangzhou 310018, P.R.China
*E-mail：happyyty@aliyun.com

**Abstract:** In this paper, by using $d$-level single-particle states, two novel multi-party quantum private comparison protocols for size relation comparison with two semi-honest third parties and one semi-honest third party are constructed, respectively. Here, each protocol can compare the size relation of secret integers from $n$ parties rather than just the equality within one time execution. In each protocol, every third party is assumed to be semi-honest in the sense that she may misbehave on her own but is not allowed to collude with anyone else; and each party employs the qudit shifting operation to encode her secret integer. Each protocol can resist both the outside attack and the participant attack. Specially, each party's secret integer can be kept unknown to other parties and the third parties. The proposed protocol with two third parties is workable in a stranger environment, as there are no communication and no pre-shared key between each pair of party. The proposed protocol with one third party is workable in an acquaintance environment, as all parties need to share a common private key beforehand.

**Keywords:** Quantum cryptography; Multi-party quantum private comparison; Size relation comparison; Semi-honest third party; $d$-level single-particle state

## 1 Introduction

Quantum cryptography can be regarded as the combination of quantum mechanics and classical cryptography. Depending on the fundamental principles of quantum mechanics, quantum cryptography has the prominent advantage of security over classical cryptography whose security relies on the computational complexity of solving mathematical problems. Since the first quantum cryptographic protocol was proposed by Bennett and Brassard in 1984 [1], many quantum cryptographic protocols have been designed to accomplish various cryptographic tasks that are intractable with classical cryptography, such as quantum key distribution (QKD) [1-7], quantum secure direct communication (QSDC) [8-10], quantum secret sharing (QSS) [11-21], quantum teleportation [22-24], quantum private query [25-28] and so on.

Secure multi-party computation (SMC), aims to calculate a function with private inputs from different parties in a distributed network on the basis of not disclosing the genuine content of each private input, and has been deeply studied in classical cryptography so far. SMC can be applied into many circumstances including private comparison, anonymous bidding and auctions, secret ballot elections, e-commerce and date mining *etc*. As an important branch of SMC, private comparison, was first introduced by Yao [29] in the millionaires' problem. In the millionaires' problem, two millionaires want to judge who is richer without knowing each other's actual property. Afterward, Boudot *et al.* [30] suggested a private comparison protocol to determine whether two millionaires are equally rich or not. However, the security of classical private comparison protocols relies on the computation complexity of solving mathematical problems, which is vulnerable to the strong ability of parallel quantum computation.

As the generalization of classical private comparison into the realm of quantum mechanics, quantum private comparison (QPC), was first suggested by Yang and Wen [31] in 2009. In recent years, the design and analysis of QPC protocols have successfully attracted much attention. Many QPC protocols [31-48] have been proposed. Since it is impossible to construct a secure equality function in a two-party scenario [49], all previous QPC protocols [31-48] have drawn support from a third party (TP). However, all of these protocols [31-48] can only compare whether two parties' secrets are equal or not instead of their size relation. Fortunately, in 2012, Lin *et al.*[50] successfully proposed a QPC protocol for comparing the size relation by using $d$-level Bell states. In 2013, Guo *et al.*[51] successfully put forward a QPC protocol for comparing the size relation based on entanglement swapping of $d$-level Bell states. In 2013, Yu *et al.*[52] successfully put forward a QPC protocol for comparing the size relation by using $d$-level single-particle states.

The protocols described above can only compare the secrets for just two parties. In 2013,

Chang et al.[53] put forward the first multi-party quantum private comparison (MQPC) protocol by using multi-particle GHZ class states, which can accomplish arbitrary pair's comparison of equality among $n$ parties within one time execution of the protocol. Since then, MQPC has attracted more and more attention so that many MQPC protocols within the multi-level system [54-59] have been designed. For example, in 2014, Luo et al.[57] proposed a MQPC protocol for comparing the size relation based on $d$-dimensional entangled states; in 2015, Huang et al.[58] proposed a MQPC protocol, in which $n$ parties can accomplish the comparison of size relation with the help of an almost-dishonest TP within one time execution of the protocol; and in 2017, Hung et al.[59] proposed a MQPC protocol for strangers to accomplish the equality comparison with the help of two almost-dishonest TPs within one time execution of the protocol.

Based on the above analysis, in this paper, by using $d$-level single-particle states, we propose two novel MQPC protocols for size relation comparison with two semi-honest TPs and one semi-honest TP, respectively. In each protocol, $n$ parties can compare the size relation of their secret integers within one time execution of the protocol. Here, each TP is semi-honest in the sense that she may misbehave on her own but is not allowed to collude with anyone else. The proposed MQPC protocol with two TPs is suitable for strangers to accomplish the size relation comparison, as there are no requirements of communication and pre-shared key between each pair of party. The proposed MQPC protocol with one TP is suitable for acquaintances to accomplish the size relation comparison, as all parties need to share a common private key beforehand.

The rest of this paper is organized as follows. In Sect.2, we present a MQPC protocol for comparing the size relation of secret integers from different parties with the help of two semi-honest TPs and analyze its security against several famous attacks; in Sect.3, we present a MQPC protocol for comparing the size relation of secret integers from different parties with the help of one semi-honest TP and analyze its security against several famous attacks; and finally, discussion and conclusion are given in Sect.4.

## 2 The proposed MQPC protocol for size relation comparison with two semi-honest TPs

In a $d$-level quantum system, a common basis of single photons can be described as follows:
$$T_1 = \{|j\rangle\}, \quad j \in \{0,1,\ldots,d-1\}. \tag{1}$$

Each element in the set $T_1$ is orthogonal to the others. After performing the $d$th order discrete quantum Fourier transform $F$ on each state in $T_1$, another basis shown in Eq.(2) forms:
$$T_2 = \{F|j\rangle\} = \left\{ \frac{1}{\sqrt{d}} \sum_{k=0}^{d-1} e^{\frac{2\pi ijk}{d}} |k\rangle \right\}, \quad j \in \{0,1,\ldots,d-1\}. \tag{2}$$

Each element in the set $T_2$ is also orthogonal to the others. The two sets, $T_1$ and $T_2$, are two common conjugate bases. The unitary operations shown in Eq.(3), represent the qudit shifting operations, where $m \in \{0,1,\ldots,d-1\}$.
$$U_m = \sum_{k=0}^{d-1} |k+m\rangle\langle k|. \tag{3}$$

### 2.1 Protocol description

Assume that the party $P_i$ has the secret integer $s_i$, where $i=1,2,\ldots,n$. There is a natural number $r$ satisfying $r > \{s_i\}_{\max}$ and $d \geq 2r-1$. The $n$ parties execute the following steps to complete the aim of size relation comparison with the help of two semi-honest TPs, $TP_1$ and $TP_2$. $TP_1$ ($TP_2$) is semi-honest in the sense that she may misbehave on her own but is not allowed to conspire with anyone else. Note that this kind of semi-honest TP belongs to the second semi-honest model of TP in Ref.[40]. Moreover, the term "semi-honest" here is equivalent to the term "almost-dishonest" in Ref.[58]. It should be emphasized that to make the protocol effective, both $TP_1$ and $TP_2$ should authenticate the identity of $P_i$ before implementing the protocol. As a result, a key which can represent the identify of $P_i$ should be pre-shared between $P_i$ and $TP_1$ ($TP_2$).

**Step 1:** $TP_1$ prepares $n$ single particles, $|k_1\rangle, |k_2\rangle, \ldots, |k_n\rangle$, which are randomly chosen from the set $T_1$. It should be ensured that $k_i$ meets the relation of $k_i + k_i' = K$. Here, $k_i \in \{0,1,\ldots,r-1\}$, $k_i' \in \{0,1,\ldots,d-1\}$, $i = 1,2,\ldots,n$ and $K$ is a constant. Note that only $TP_1$ knows the genuine value of $K$.

**Step 2:** $TP_1$ prepares $n$ groups of $l$ decoy photons, which are denoted as $G_1, G_2, \ldots, G_n$, respectively. Here, each decoy photon is one of the states randomly chosen from the set $T_1$ or $T_2$. Afterward, $TP_1$ inserts the particle $|k_i\rangle$ into $G_i$ at random positions to construct a new sequence $G_i'$. Finally, $TP_1$ sends $G_i'$ to $P_i$, where $i = 1,2,\ldots,n$.

**Step 3:** After confirming that $P_i$ ($i = 1,2,\ldots,n$) has received all particles, $TP_1$ and $P_i$ check the security of the transmission of $G_i'$. Concretely, $TP_1$ announces the positions and the preparation bases of decoy photons to $P_i$. According to the announced information, $P_i$ uses the right bases to measure the corresponding decoy photons and returns the measurement results to $TP_1$. Afterward, by comparing the measurement results of decoy photons with their corresponding initial prepared states, $TP_1$ can check whether the transmission of $G_i'$ is secure or not. If it is secure, the protocol will be proceeded to the next step; otherwise, the protocol will be terminated and restarted from Step 1.

**Step 4:** After removing the decoy photons in $G_i'$ ($i = 1,2,\ldots,n$), in order to encode her secret integer $s_i$, $P_i$ performs the operation $U_{s_i} = \sum_{k=0}^{d-1} |k+s_i\rangle\langle k|$ on the particle $|k_i\rangle$ to obtain the state $|k_i + s_i\rangle$. It should be pointed out that $k_i + s_i < d$. Afterward, $P_i$ prepares a quantum state sequence $G_i''$ composed of $l$ decoy photons. Here, each decoy photon is also one of the states randomly chosen from the set $T_1$ or $T_2$. Then, $P_i$ inserts the particle $|k_i + s_i\rangle$ into $G_i''$ at random positions to construct a new sequence $G_i'''$. Finally, $P_i$ sends $G_i'''$ to $TP_2$.

**Step 5:** After confirming that $TP_2$ has received all particles, $P_i$ ($i = 1,2,\ldots,n$) and $TP_2$ check the security of the transmission of $G_i'''$. Firstly, $P_i$ announces the positions and the preparation bases of decoy photons which are chosen from the set $T_2$. According to the announced information, $TP_2$ uses the right bases to measure the corresponding decoy photons and returns the measurement results to $P_i$. Then, by comparing the measurement results of decoy photons with their corresponding initial prepared states, $P_i$ can calculate the error rate. If the error rate is greater than the threshold value, the protocol will be terminated and restarted from Step 1; otherwise, the protocol will be proceeded to the next step.

**Step 6:** $TP_2$ drops out the decoy photons in $G_i'''$ ($i = 1,2,\ldots,n$) which are chosen from the set $T_2$. For the remaining particles on the site of $TP_2$, $P_i$ tells $TP_2$ the positions and the preparation bases of decoy photons which are chosen from the set $T_1$. Afterward, $TP_2$ uses the right bases to measure the corresponding decoy photons and returns the measurement results to $P_i$. Then, by comparing the measurement results of decoy photons with their corresponding initial prepared states, $P_i$ can calculate the error rate. If the error rate is greater than the threshold value, the protocol will be terminated and restarted from Step 1; otherwise, the protocol will be proceeded to the next step.

**Step 7:** $TP_2$ uses the correct base to measure the particle $|k_i + s_i\rangle$ ($i = 1,2,\ldots,n$) and obtains the value of $k_i + s_i$. Then, $TP_1$ announces $k_1', k_2', \ldots, k_n'$ to $TP_2$ through the classical channel. Afterward, $TP_2$ calculates the sum of $k_i + s_i$ and $k_i'$, i.e., $M_i = k_i + s_i + k_i'$. Because $k_i + k_i' = K$, the size relation of different $M_i$ s is the same to that of different $s_i$ s. As a result, $TP_2$ can know the size relation of different $s_i$ s, but cannot know the genuine value of $s_i$ due to lack of knowledge about $k_i$. Finally, $TP_2$ announces the size relation of different $M_i$ s through the classical channel. According to the announced information, $P_i$ can know the size relation of different $s_i$ s.

Now it concludes the description of the proposed MQPC protocol with two TPs. For clarity, we further show its flow chart in Fig.1, taking $TP_1$, $P_i$ and $TP_2$ for example, after ignoring the security check processes.

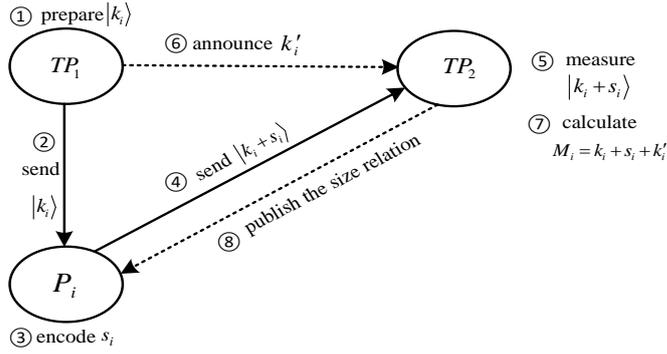

Fig.1 The flow chart of the proposed MQPC protocol with two TPs

(Here, we ignore the security check processes and take $TP_1$, $P_i$ and $TP_2$ for example. The solid line with an arrow denotes the quantum state transmission operation while the dotted line with an arrow represents the classical information transmission operation)

## 2.2 Security analysis

In this subsection, we analyze the security of the proposed MQPC protocol with two TPs. Firstly, we show that the outside attack is invalid to the proposed MQPC protocol with two TPs. Secondly, we show that one or more parties cannot learn other parties' secret integers. Moreover, the semi-honest $TP_1$ and $TP_2$ cannot learn these parties' secret integers either.

**A  The outside attack**

We analyze the possibility for an outside eavesdropper to steal the parties' secret integers according to each step of the proposed MQPC protocol.

In Step 1, there is nothing transmitted. Hence, an outside eavesdropper has no chance to launch her attacks in this step.

In Step 2, $TP_1$ sends $G'_i$ to $P_i$; and in Step 4, $P_i$ sends $G'''_i$ to $TP_2$. Here, $i = 1,2,\ldots,n$. An outside eavesdropper may utilize the qudit transmissions in these two steps to extract some useful information about the parties' secret integers. Obviously, both of these two steps use the decoy photon technique [60,61] to guarantee the security of qudit transmissions. The decoy photon technology can be thought as a variant of the eavesdropping check method of the BB84 protocol [1] which has been proven to be unconditionally secure [62]. The effectiveness of decoy photon technology for 2-level quantum system towards the intercept-resend attack, the measure-resend attack and the entangle-measure attack has also been validated in Refs.[63,64]. It is straightforward that the decoy photon technology for $d$-level quantum system is also effective towards these famous attacks. Therefore, an outside eavesdropper cannot extract any useful information about secret integers without being detected by the eavesdropping check processes in Steps 3, 5 and 6.

In Step 7, $TP_1$ announces $k'_1, k'_2, \ldots, k'_n$ to $TP_2$ and $TP_2$ announces the size relation of different $M_i$s ($i = 1,2,\ldots,n$) through the classical channel. Even though an outside eavesdropper may hear of $k'_1, k'_2, \ldots, k'_n$ and the comparison result of size relation, these information is still helpless for her to obtain the genuine values of secret integers.

In addition, it should be emphasized that in order to overcome the invisible photon eavesdropping Trojan horse attack [65], $P_i$ should insert a filter in front of her devices to filter out the photon signal with an illegitimate wavelength [66,67]. Moreover, in order to resist the delay-photon Trojan horse attack [66,68], $P_i$ should adopt a photon number splitter (PNS:50/50) to split each sample quantum signal into two pieces and measure the signals after the PNS with proper measuring bases [66,67]. If the multiphoton rate is unreasonably great, this attack will be discovered.

**B  The participant attack**

In 2007, Gao *et al*. [69] first pointed out that the attack from a dishonest participant is generally more powerful and should be paid more attention to. Up to now, the participant attack has attracted much attention in the cryptanalysis of quantum cryptography [70-72]. In this subsection, we consider three cases of participant attack. Firstly, we discuss the attack from one or more dishonest parties; secondly, we discuss the participant attack from semi-honest $TP_1$; and thirdly, we discuss the participant attack from semi-honest $TP_2$.

**Case 1: The participant attack from one or more dishonest parties**

Two situations should be considered in this case. One is that one dishonest party wants to steal other parties' secret integers; and the other one is that two or more dishonest parties conclude together to steal other parties' secret integers.

**(a) The participant attack from one dishonest party**

It is apparent that in the proposed MQPC protocol with two TPs, the roles of $n$ parties are the same. Without loss of generality, in this situation, we suppose that $P_1$ is the only dishonest party who wants to steal other parties' secret integers.

In the proposed MQPC protocol with two TPs, there are qudits transmitted from $TP_1$ to $P_j$ ($j = 2,3,\ldots,n$) in Step 2 and from $P_j$ to $TP_2$ in Step 4. However, there is not any qudit transmitted between $P_1$ and $P_j$. As a result, $P_1$ is thoroughly independent from $P_j$. In order to obtain the secret integer of $P_j$, when $P_1$ tries to launch her attacks on the transmitted particles from $TP_1$ to $P_j$ in Step 2 or from $P_j$ to $TP_2$ in Step 4, she essentially acts as an outside eavesdropper. As analyzed in part A of Sect.2.2, she will inevitably be caught if she launches her attacks, as she has no knowledge about the positions and the preparation bases of transmitted decoy photons.

In addition, in Step 7, $P_1$ may hear of $k'_1, k'_2, \ldots, k'_n$ from $TP_1$ and the comparison result of size relation from $TP_2$. However, this information is still helpless for her to obtain the secret integer of $P_j$.

It can be concluded now that one dishonest party cannot obtain other parties' secret integers.

**(b) The participant attack from two or more dishonest parties**

If the number of dishonest parties concluding together is $n-1$, the dishonest parties will be the most powerful. Here, we only consider this extreme case. Without loss of generality, suppose that the dishonest $P_1, P_2, \ldots, P_{m-1}, P_{m+1}, \ldots, P_n$ collude together to get the secret integer of $P_m$.

It is apparent that $P_1, P_2, \ldots, P_{m-1}, P_{m+1}, \ldots, P_n$ are thoroughly independent from $P_m$. If $P_1, P_2, \ldots, P_{m-1}, P_{m+1}, \ldots, P_n$ try to launch their attacks on the transmitted particles from $TP_1$ to $P_m$ in Step 2 or from $P_m$ to $TP_2$ in Step 4, as analyzed in part A of Sect.2.2, they will inevitably be detected as an outside eavesdropper, since they have no knowledge about the positions and the preparation bases of transmitted decoy photons.

In addition, in Step 7, $P_1, P_2, \ldots, P_{m-1}, P_{m+1}, \ldots, P_n$ may hear of $k'_1, k'_2, \ldots, k'_n$ from $TP_1$ and the comparison result of size relation from $TP_2$. However, this information is still helpless for them to obtain the secret integer of $P_m$.

It can be concluded now that the dishonest $P_1, P_2, \ldots, P_{m-1}, P_{m+1}, \ldots, P_n$ cannot get the secret integer of $P_m$. It is straightforward that the colluding attack from two or more dishonest parties is invalid for them to steal the secret integers from other parties.

**Case 2: The participant attack from semi-honest $TP_1$**

In the proposed MQPC protocol with two TPs, $TP_1$ is not allowed to conspire with anyone else including $TP_2$. As a result, $TP_1$ cannot know the state of $|k_i + s_i\rangle$ ($i = 1,2,\ldots,n$) from $TP_2$. In order to obtain the state of $|k_i + s_i\rangle$, $TP_1$ may further try to intercept the transmitted particles from $P_i$ to $TP_2$ in Step 4, measure them with the base $T_1$ and resend the measured particles to $TP_2$. However, this attack from $TP_1$ will be easily detected by the security check process in Step 5, as it alters the states of decoy photons chosen from the set $T_2$. It can be concluded that $TP_1$ has no knowledge about the state of $|k_i + s_i\rangle$ at all.

In Step 1, $TP_1$ knows the states of $|k_1\rangle, |k_2\rangle, \ldots, |k_n\rangle$ and the values of $k'_1, k'_2, \ldots, k'_n$. In Step 7, $TP_1$ may hear of the comparison result of size relation from $TP_2$. However, she still cannot infer out the secret integer of $P_i$ from this information, as she has no chance to know the state of $|k_i + s_i\rangle$.

**Case 3: The participant attack from semi-honest $TP_2$**

In the proposed MQPC protocol with two TPs, $TP_2$ is not allowed to conspire with anyone else including $TP_1$. The main task of $TP_2$ is to measure the state of $|k_i + s_i\rangle$ ($i = 1,2,\ldots,n$), calculate the sum of $k_i + s_i$ and $k_i'$, and compare the size relation of secret integers from $n$ parties. In Step 7, $TP_2$ can know the state of $|k_i + s_i\rangle$. In order to obtain the secret integer of $P_i$, $TP_2$ needs to know the state of $|k_i\rangle$. As a result, $TP_2$ may try to intercept the transmitted particles from $TP_1$ to $P_i$ in Step 2, measure them with the base $T_1$ and resend the measured particles to $P_i$. However, he will inevitably be detected by the security check process in Step 3, as this attack alters the states of decoy photons chosen from the set $T_2$. As a result, $TP_2$ has no chance to obtain the state of $|k_i\rangle$. It can be concluded that $TP_2$ cannot know the secret integer of $P_i$.

## 3 The proposed MQPC protocol for size relation comparison with one semi-honest TP

### 3.1 Protocol description

Assume that the party $P_i$ has the secret integer $s_i$, where $i = 1,2,\ldots,n$. And $n$ parties share a key (i.e. $C$) beforehand through the BB84 QKD protocol [1], where $C$ is a constant. There is a natural number $r$ satisfying $r > \{s_i\}_{\max}$, $r > C$ and $d \geq 3r - 1$. The $n$ parties execute the following steps to complete the aim of size relation comparison with the help of one semi-honest TP. Same to the protocol in Sect.2.1, TP is semi-honest also in the sense that she may misbehave on her own but is not allowed to conspire with anyone else.

**Step 1:** TP prepares $n$ single particles, $|k_1\rangle, |k_2\rangle, \ldots, |k_n\rangle$, which are randomly chosen from the set $T_1$. It should be ensured that $k_i$ meets the relation of $k_i + k_i' = K$. Here, $k_i \in \{0,1,\ldots,r-1\}$, $k_i' \in \{0,1,\ldots,d-1\}$, $i = 1,2,\ldots,n$ and $K$ is a constant. Note that only TP knows the genuine value of $K$.

**Step 2:** TP prepares $n$ groups of $l$ decoy photons, which are denoted as $G_1, G_2, \ldots, G_n$, respectively. Here, each decoy photon is one of the states randomly chosen from the set $T_1$ or $T_2$. Afterward, TP inserts the particle $|k_i\rangle$ into $G_i$ at random positions to construct a new sequence $G_i'$. Finally, TP sends $G_i'$ to $P_i$, where $i = 1,2,\ldots,n$.

**Step 3:** After confirming that $P_i$ ($i = 1,2,\ldots,n$) has received all particles, TP and $P_i$ check the security of the transmission of $G_i'$. Concretely, TP announces the positions and the preparation bases of decoy photons to $P_i$. According to the announced information, $P_i$ uses the right bases to measure the corresponding decoy photons and returns the measurement results to TP. Afterward, by comparing the measurement results of decoy photons with their corresponding initial prepared states, TP can check whether the transmission of $G_i'$ is secure or not. If it is secure, the protocol will be proceeded to the next step; otherwise, the protocol will be terminated and restarted from Step 1.

**Step 4:** After removing the decoy photons in $G_i'$ ($i = 1,2,\ldots,n$), in order to encode her secret integer $s_i$, $P_i$ performs the operation $U_{s_i+C} = \sum_{k=0}^{d-1} |k + s_i + C\rangle\langle k|$ on the particle $|k_i\rangle$ to obtain the state $|k_i + s_i + C\rangle$. It should be pointed out that $k_i + s_i + C < d$. Afterward, $P_i$ prepares a quantum state sequence $G_i''$ composed of $l$ decoy photons. Here, each decoy photon is also one of the states randomly chosen from the set $T_1$ or $T_2$. Then, $P_i$ inserts the particle $|k_i + s_i + C\rangle$ into $G_i''$ at random positions to construct a new sequence $G_i'''$. Finally, $P_i$ sends $G_i'''$ to TP.

**Step 5:** After confirming that TP has received all particles, $P_i$ ($i = 1,2,\ldots,n$) and TP check the security of the transmission of $G_i'''$. Firstly, $P_i$ announces the positions and the preparation bases of decoy photons which are chosen from the set $T_2$. According to the announced information, TP uses the right bases to measure the corresponding decoy photons and returns the measurement results to $P_i$. Then, by comparing the measurement results of decoy photons with their corresponding initial prepared states, $P_i$ can calculate the error rate. If the error rate is greater than the threshold value, the protocol will be terminated and restarted from Step 1; otherwise, the protocol will be proceeded to the next step.

**Step 6:** TP drops out the decoy photons in $G_i'''$ ($i=1,2,\ldots,n$) which are chosen from the set $T_2$. For the remaining particles on the site of TP, $P_i$ tells TP the positions and the preparation bases of decoy photons which are chosen from the set $T_1$. Afterward, TP uses the right bases to measure the corresponding decoy photons and returns the measurement results to $P_i$. Then, by comparing the measurement results of decoy photons with their corresponding initial prepared states, $P_i$ can calculate the error rate. If the error rate is greater than the threshold value, the protocol will be terminated and restarted from Step 1; otherwise, the protocol will be proceeded to the next step.

**Step 7:** TP uses the correct base to measure the particle $|k_i+s_i+C\rangle$ ($i=1,2,\ldots,n$) and obtains the value of $k_i+s_i+C$. Then, TP calculates the the sum of $k_i+s_i+C$ and $k_i'$, i.e., $M_i=k_i+s_i+C+k_i'$. Because both $k_i+k_i'=K$ and $C$ are constant, the size relation of different $M_i$s is the same to that of different $s_i$s. As a result, TP can know the size relation of different $s_i$s, but cannot know the genuine value of $s_i$ due to lack of knowledge about $C$. Finally, TP announces the size relation of different $M_i$s through the classical channel. According to the announced information, $P_i$ can know the size relation of different $s_i$s.

Now it concludes the description of the proposed MQPC protocol with one TP. For clarity, we further show its flow chart in Fig.2, taking TP and $P_i$ for example, after ignoring the security check processes.

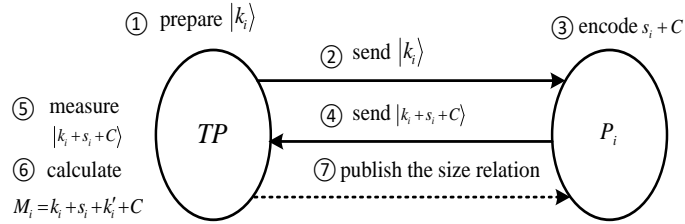

Fig.2 The flow chart of the proposed MQPC protocol with one TP
(Here, we ignore the security check processes and take TP and $P_i$ for example)

## 3.2 Security analysis

In this subsection, we analyze the security of the proposed MQPC protocol with one TP by concentrating on the participant attack, since its security analysis on the outside attack is extremely similar to that of the proposed protocol with two TPs. In the following, we show that one or more parties cannot learn other parties' secret integers. Moreover, the semi-honest TP cannot learn these parties' secrets either.

**Case 1: The participant attack from one or more dishonest parties**

Two situations should also be considered in this case. One is that one dishonest party wants to steal other parties' secret integers; and the other one is that two or more dishonest parties conclude together to steal other parties' secret integers.

**(a) The participant attack from one dishonest party**

It is apparent that in the proposed MQPC protocol with one TP, the roles of $n$ parties are the same. Without loss of generality, in this situation, we suppose that $P_1$ is the only dishonest party who wants to steal other parties' secret integers.

In the proposed MQPC protocol with one TP, there are qudits transmitted from TP

to $P_j$ ( $j = 2,3,\ldots,n$ ) in Step 2 and from $P_j$ to TP in Step 4. In order to obtain the secret integer of $P_j$, when $P_1$ tries to launch her attacks on the transmitted particles from TP to $P_j$ in Step 2 or from $P_j$ to TP in Step 4, she essentially acts as an outside eavesdropper. As analyzed in part A of Sect.2.2, she will inevitably be caught if she launches her attacks, as she has no knowledge about the positions and the preparation bases of transmitted decoy photons.

In addition, in Step 7, $P_1$ may hear of the comparison result of size relation from TP. However, this information is still helpless for her to obtain the secret integer of $P_j$.

It can be concluded now that one dishonest party cannot obtain other parties' secret integers.

**(b) The participant attack from two or more dishonest parties**

If the number of dishonest parties concluding together is $n-1$, the dishonest parties will be the most powerful. Here, we only consider this extreme case. Without loss of generality, suppose that the dishonest $P_1, P_2, \ldots, P_{m-1}, P_{m+1}, \ldots, P_n$ collude together to get the secret integer of $P_m$.

If $P_1, P_2, \ldots, P_{m-1}, P_{m+1}, \ldots, P_n$ try to launch their attacks on the transmitted particles from TP to $P_m$ in Step 2 or from $P_m$ to TP in Step 4, as analyzed in part A of Sect.2.2, they will inevitably be detected as an outside eavesdropper, since they have no knowledge about the positions and the preparation bases of transmitted decoy photons.

In addition, in Step 7, $P_1, P_2, \ldots, P_{m-1}, P_{m+1}, \ldots, P_n$ may hear of the comparison result of size relation from TP. However, this information is still helpless for them to obtain the secret integer of $P_m$.

It can be concluded now that the dishonest $P_1, P_2, \ldots, P_{m-1}, P_{m+1}, \ldots, P_n$ cannot get the secret integer of $P_m$. It is straightforward that the colluding attack from two or more dishonest parties is invalid for them to steal the secret integers from other parties.

**Case 2: The participant attack from semi-honest TP**

In Step 1, TP knows the states of $|k_1\rangle, |k_2\rangle, \ldots, |k_n\rangle$ and the values of $k_1', k_2', \ldots, k_n'$. In Step 7, TP obtains $k_i + s_i + C$ ( $i = 1,2,\ldots,n$ ) by measuring the state of $|k_i + s_i + C\rangle$, calculates the sum of $k_i + s_i + C$ and $k_i'$, and compares the size relation of secret integers from $n$ parties. However, she still cannot infer out the secret integer of $P_i$ from this information, as she has no chance to know the pre-shared key (i.e. $C$ ) of $P_i$.

## 4 Discussion and conclusion

Firstly, we compare the proposed MQPC protocols with the QPC protocols of Refs.[50-52] without considering the security check processes. Table 1 clearly shows that the performance of the proposed MQPC protocols has advantages over the protocols of Refs.[50-52]. For example, in the protocols of Refs.[50-52], in order to complete the whole size relation comparisons for $n$ parties, TP has to execute the same protocol $x$ times, where $x = (n-1)$ to $\frac{n(n-1)}{2}$. Accordingly, in the optimal case, TP can complete the whole size relation comparisons within $n-1$ times execution; and in the worst case, TP has to execute the protocol $\frac{n(n-1)}{2}$ times. However, the proposed MQPC protocols can complete the whole size relation comparisons for $n$ parties within one execution.

Secondly, we compare the proposed MQPC protocols with the MQPC protocol of Refs.[53-58] without considering the security check processes. Note that there are two MQPC protocols in Ref.[55], which are denoted as Ref.[55]-A and Ref.[55]-B, respectively. In the protocols of Refs.[53-56], TP can only compare all parties' secrets for equality while in the protocols of Refs.[57-58] and the proposed MQPC protocols, TP can compare the size relation of all parties' secrets. In addition, the proposed MQPC protocols just need $d$ -level single-particle states as quantum resource while all of other MQPC protocols listed in Table 2 employ $d$ -level entangled states as quantum resource. As the preparation of $d$ -level single-particle states is much easier than that of $d$ -level entangled states, the proposed MQPC protocols exceed other MQPC protocols listed in Table 2 on the aspect of quantum resource.

Table 1  Comparison between the proposed MQPC protocols and the QPC protocols of Refs.[50-52]

| | The protocol of Ref.[50] | The protocol of Ref.[51] | The protocol of Ref.[52] | The proposed protocol with two TPs | The proposed protocol with one TP |
|---|---|---|---|---|---|
| Quantum state | $d$-level two-particle Bell state | $d$-level two-particle Bell state | $d$-level single-particle state | $d$-level single-particle state | $d$-level single-particle state |
| Number of parties | 2 | 2 | 2 | $n$ | $n$ |
| Pre-shared QKD key | Yes | Yes | Yes | Yes | Yes |
| Number of times to compare $n$ parties | $n-1 \sim \frac{n(n-1)}{2}$ | $n-1 \sim \frac{n(n-1)}{2}$ | $n-1 \sim \frac{n(n-1)}{2}$ | 1 | 1 |
| Type of TP | The second semi-honest model of TP | The second semi-honest model of TP | The second semi-honest model of TP | The second semi-honest model of TP | The second semi-honest model of TP |
| Quantum measurement for TP | No | $d$-level two-particle Bell state measurement | No | $d$-level single-particle measurement | $d$-level single-particle measurement |
| Quantum measurement for parties | $d$-level single-particle measurement | $d$-level two-particle Bell state measurement | $d$-level single-particle measurement | No | No |
| Comparison of size relation | Yes | Yes | Yes | Yes | Yes |
| Quantum technology used | Quantum fourier transform and unitary operation | Quantum entanglement swapping and unitary operation | Quantum fourier transform and unitary operation | Unitary operation | Unitary operation |

In summary, in this paper, by using $d$-level single-particle states, we propose two novel MQPC protocols for size relation comparison with two semi-honest TPs and one semi-honest TP respectively, where each TP may misbehave on her own but is not allowed to collude with anyone else. Each of the proposed MQPC protocol can compare the size relation of secret integers from $n$ parties rather than just the equality within one time execution of the protocol. We validate in detail that the proposed MQPC protocols can resist both the outside attack and the participant attack. Specially, each party has no chance to know other parties' secret integers, and the TPs also have no knowledge about each party's secret integer. As there are no communication and no pre-shared key between each pair of party, the proposed MQPC protocol with two TPs is workable in a stranger environment, such as anonymous bidding and auction, where any two bidders are strange or want to keep their identities secret to each other. The proposed MQPC protocol with one TP is workable in an acquaintance environment, as all parties need to share a common private key beforehand.

## Acknowledgments

The authors would like to thank the anonymous reviewers for their valuable comments that help enhancing the quality of this paper. Funding by the National Natural Science Foundation of China (Grant No.61402407) and the Natural Science Foundation of Zhejiang Province (Grant No.LY18F020007) is gratefully acknowledged.

Table 2  Comparison between the proposed MQPC protocols and the MQPC protocols of Refs.[53-58]

| | Quantum state | Quantum measurement for TP | Quantum measurement for parties | Number of parties | Quantum technology used | Type of TP | Comparison of size relation | Pre-shared QKD key | Number of times to compare $n$ parties |
|---|---|---|---|---|---|---|---|---|---|
| The protocol of Ref.[53] | $n$-particle GHZ class state | No | single-particle measurement | $n$ | The entanglement correlation among different particles from one quantum entangled state | The first kind of semi-honest TP | No | No | 1 |
| The protocol of Ref.[54] | $d$-level $n$-particle entangled state | $d$-level single-particle measurement | No | $n$ | Quantum fourier transform and unitary operation | The first kind of semi-honest TP | No | No | 1 |
| The protocol of Ref.[55]-A | $d$-level $n$-particle entangled state and $d$-level two-particle entangled | $d$-level single-particle measurement | $d$-level single-particle measurement | $n$ | Quantum fourier transform | The second semi-honest model of TP | No | No | 1 |
| | state | | | | | | | | |
| The protocol of Ref.[55]-B | $d$-level two-particle entangled state | $d$-level two-particle collective measurement | No | $n$ | Unitary operation | The second semi-honest model of TP | No | Yes | 1 |
| The protocol of Ref.[56] | $d$-level $n+1$-particle cat state and $d$-level two-particle Bell state | $d$-level $n+1$-particle cat state measurement | $d$-level two-particle Bell state measurement | $n$ | Quantum entanglement swapping and unitary operation | The second semi-honest model of TP | No | No | 1 |
| The protocol of Ref.[57] | $d$-level $l$-particle entangled state | No | $d$-level single-particle measurement | $n$ | The entanglement correlation among different particles from one quantum entangled state | The second semi-honest model of TP | Yes | Yes | 1 |
| The protocol of Ref.[58] | $d$-level $n$-particle GHZ state and $l$-level $n$-particle GHZ state | $d$-level single-particle measurement | $d$-level single-particle measurement | $n$ | Unitary operation | The second semi-honest model of TP | Yes | No | 1 |
| The proposed MQPC protocol with two TPs | $d$-level single-particle state | $d$-level single-particle measurement | No | $n$ | Unitary operation | The second semi-honest model of TP | Yes | Yes | 1 |
| The proposed MQPC protocol with one TP | $d$-level single-particle state | $d$-level single-particle measurement | No | $n$ | Unitary operation | The second semi-honest model of TP | Yes | Yes | 1 |